# Higher-order mode-based cavity misalignment measurements at the free-electron laser FLASH

Thorsten Hellert,[*] Nicoleta Baboi, and Liangliang Shi

*DESY, Notkestrasse 85, 22603 Hamburg, Germany*

(Received 19 August 2017; published 19 December 2017)

At the Free-Electron Laser in Hamburg (FLASH) and the European X-Ray Free-Electron Laser, superconducting TeV-energy superconducting linear accelerator (TESLA)-type cavities are used for the acceleration of electron bunches, generating intense free-electron laser (FEL) beams. A long rf pulse structure allows one to accelerate long bunch trains, which considerably increases the efficiency of the machine. However, intrabunch-train variations of rf parameters and misalignments of rf structures induce significant trajectory variations that may decrease the FEL performance. The accelerating cavities are housed inside cryomodules, which restricts the ability for direct alignment measurements. In order to determine the transverse cavity position, we use a method based on beam-excited dipole modes in the cavities. We have developed an efficient measurement and signal processing routine and present its application to multiple accelerating modules at FLASH. The measured rms cavity offset agrees with the specification of the TESLA modules. For the first time, the tilt of a TESLA cavity inside a cryomodule is measured. The preliminary result agrees well with the ratio between the offset and angle dependence of the dipole mode which we calculated with eigenmode simulations.

DOI: 10.1103/PhysRevAccelBeams.20.123501

## I. INTRODUCTION

The Free-Electron Laser in Hamburg (FLASH) [1,2] and European X-Ray Free-Electron Laser (European XFEL) [3–5] are single pass free-electron lasers (FELs), generating high-brilliance radiation by self-amplified spontaneous emission [6]. Acceleration of the driving electron bunches is achieved by using superconducting TeV-energy superconducting linear accelerator (TESLA) [7] -type cavities. The high duty cycle and, thus, long radio frequency (rf) pulse structure allows one to provide long bunch trains adapted to the needs of the experiments, which significantly increases the efficiency of the machine. High longitudinal and transverse stability of the beam along the bunch train is essential for multibunch FEL operation. However, intrabunch-train variations of rf parameters as well as structure misalignments induce intrabunch-train trajectory variations which affect the multibunch FEL performance considerably [8]. The accelerating cavities are housed inside 12-m-long cryomodules [9], which restricts the ability for direct alignment measurements. Precise knowledge of the cavity misalignment, however, would allow for optimizing the low-level rf in order to compensate the impact on the intrabunch-train trajectory variation.

In order to determine the transverse cavity position, we chose a method based on beam-excited dipole modes in the cavities. These modes have a linear dependence on the transverse beam offset and angle with respect to the cavity axis. When the beam trajectory is varied systematically, the relative strength of the excited modes can be compared quantitatively, and the determination of the cavity axis becomes a linear regression problem. The proof of principle measurements can be found in Ref. [10]. Because of the low beam energy, and thus beam sensitivity to off-axis fields, the injector module (ACC1) has the largest impact on the intrabunch-train trajectory variation, and therefore knowledge on its cavity alignment is the first priority. Dedicated studies to derive accelerating structure misalignments from multibunch rf and beam position measurements [11] at the injector module have not been conclusive.

We have developed a measurement and signal processing routine which is able to reliably evaluate a large amount of data. Its application to multiple accelerating modules at FLASH will be presented. We will start this paper with a brief summary of the principles of dipole modes needed for the measurements. A description of the signal measurement and processing procedure follows. Because of the large size of the accelerating cavities, ambiguities arise in terms of the beam position when the beam travels obliquely with respect to the cavity axis. For the first time, we quantified this effect experimentally. We compare our results with simulations based on eigenmode calculations of a TESLA cavity. Finally, the results of cavity misalignment measurements at five accelerating modules at FLASH are presented. The measurement at the injector module was the most challenging part of the experiment.

[*]thorsten.hellert@desy.de

## II. MEASUREMENT PRINCIPLES

When a bunch of electrons traverse a cavity, wakefields covering a wide range in frequency domains are excited. These fields can be classified into different modes according to the field distribution [12]. The modes with higher frequencies than the fundamental mode used for acceleration are referred to as higher-order modes (HOMs). These modes can damage the beam quality, and therefore they are damped. For this reason, each TESLA cavity is equipped with two HOM couplers [13]. It has been shown experimentally [10] that a beam-excited dipole mode, extracted by HOM couplers, can be used to study the misalignment of the cavities in an installed cryomodule. In this paper, we restrict the description to the basics of dipole modes and the properties needed for the HOM-based cavity misalignment measurement and its data analysis.

In cylindrically symmetric cavities, HOMs can be characterized by their azimuthal dependence as monopole, dipole, quadrupole, etc., modes [12]. According to the nine-cell geometry of the TESLA cavity, nine modes form a band. In this paper, we focus on the sixth transverse electric dipole mode, $TE_{111}$-6.

The main reason for using this mode is that the electronics installed at the first five cryomodules in FLASH filter this mode. This is one of the strongest dipole modes, quantified by an $(R/Q)$ of 5.5 $\Omega/cm^2$ [13]. One can argue that mode $TE_{111}$-7 may give better accuracy due to its higher $(R/Q)$ of 7.7 $\Omega/cm^2$. However, there are other factors that contribute to the final result, e.g., the amount of the mode energy that is extracted through the HOM coupler. Also, looking at the longitudinal electrical field distribution, the middle cell seems to contribute more to the sixth than to the seventh mode, but this remains to be shown.

Dipole modes can be excited by a traversing beam in three different scenarios, which are illustrated in Fig. 1: The trajectory may have an offset $x$ with respect to the cavity axis (top) and a tilt angle $\alpha$ (center). Additionally, the bunch itself might be tilted by an angle $\Theta$ (bottom). It has been shown [14] that the corresponding amplitude of the beam-excited dipole mode is proportional to the bunch charge $Q$ and

$$V_x(t) \propto x \cdot e^{-(t/2\tau)} \sin(\omega t), \quad (1)$$

$$V_\alpha(t) \propto -\alpha \cdot e^{-(t/2\tau)} \cos(\omega t), \quad (2)$$

$$V_\Theta(t) \propto \Theta \cdot e^{-(t/2\tau)} \cos(\omega t), \quad (3)$$

with the frequency $\omega$ and decay time $\tau$ of the considered dipole mode. The signal excited by the tilt of the bunch, $V_\Theta(t)$, is furthermore proportional to the square of the bunch length. At FLASH and European XFEL, the bunch lengths are on the order of 100 fs. $V_\Theta(t)$ is therefore vanishing small and can be neglected [10].

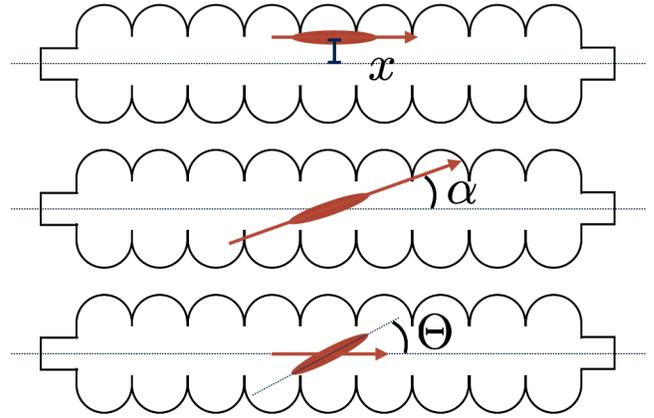

FIG. 1. Different scenarios for a bunch (red) traversing a cavity. In the top row, the paraxial trajectory has an offset $x$ with respect to the cavity axis, and the middle row shows a trajectory tilt angle $\alpha$. The bottom scenario illustrates a tilted bunch with angle $\Theta$ traversing the cavity on axis.

It is worth noting that, due to field disturbances caused by structure imperfections and couplers, the geometrical axis of the cavity can deviate slightly from the electrical axis of the considered dipole mode. This effect is hardly quantifiable, and eventually beam offsets from the electrical axis of the strongest HOMs affect the beam much more than from the axis defined by the geometry. The following analysis will therefore not distinguish between these two axes and refer to them as "cavity axis".

Dipole modes occur in orthogonally polarized doublets, following from the two transverse degrees of freedom in an axially symmetric cavity. Note that their polarization axes may not be coincident with the horizontal and vertical planes of the cavity. Furthermore, due to asymmetries and imperfections of the cavity, their frequencies are usually split, and the angle between their polarization axes can deviate from 90°. The TESLA cavities at FLASH are equipped with a HOM coupler at both ends which span an angle of about 115°. An illustration can be found in Figs. 2 and 3. Hence, depending on the particular imperfections of one cavity, the two couplers are expected to have a dissimilar sensitivity to the dipole doublet.

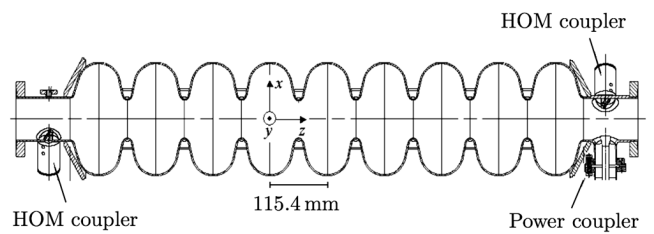

FIG. 2. Longitudinal cross section of a TESLA cavity.

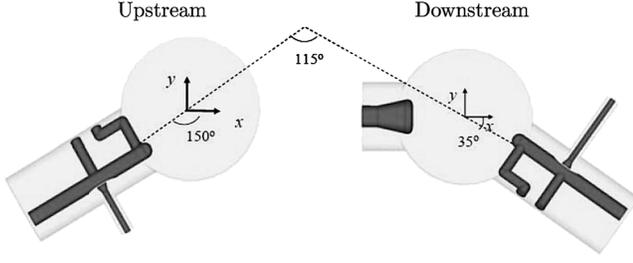

FIG. 3. Geometry and orientation of the higher-order mode (upstream and downstream) and fundamental power coupler (downstream).

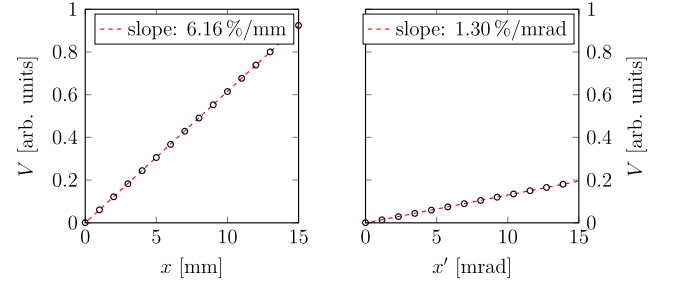

FIG. 4. Normalized amplitude $V$ of the beam-excited $TE_{111}$-6 dipole mode for different beam trajectories. The dots are results of eigenmode calculations, and the dashed lines indicate a linear fit. In the left graph, paraxial trajectories with different offsets $x$ are evaluated; the right graph shows the amplitude as calculated for different trajectory tilt angles $x'$ through the longitudinal center of the cavity.

## III. EIGENMODE SIMULATIONS

The scaling factor between the dipole mode amplitude excited by the bunch trajectory offset and the amplitude from the trajectory tilt angle depends on the mode phase change per cell and the geometry of the cavity. Previous estimations for a single cell imply that the amount of amplitude induced by an offset of 100 $\mu$m is equivalent to an amplitude excited by a tilt angle of 1 mrad [15]. In order to determine the proper ratio for the TESLA cavity, eigenmode calculations are performed [16].

The electric and magnetic field in a cavity, $\mathbf{E}$ and $\mathbf{H}$, respectively, can be expanded in terms of orthogonal eigenfunctions $\mathbf{e}^{(m)}$ and $\mathbf{h}^{(m)}$:

$$\mathbf{E}(x,y,z,t) = \mathrm{Re}\left\{\sum_m q^{(m)}(t) \cdot \mathbf{e}^{(m)}(x,y,z)\right\}, \quad (4)$$

$$\mathbf{H}(x,y,z,t) = \mathrm{Re}\left\{\sum_m p^{(m)}(t) \cdot \mathbf{h}^{(m)}(x,y,z)\right\} \quad (5)$$

with the time-dependent eigenmode amplitudes $q^{(m)}$ and $p^{(m)}$ of mode $m$. The eigenfunctions $\mathbf{e}^{(m)}$ and $\mathbf{h}^{(m)}$ are the eigenmode field distribution normalized by the energy stored in the mode. By inserting Eqs. (4) and (5) into the Maxwell equations in a vacuum, the second-order differential equation for the amplitude of the electric field can be obtained as

$$\frac{d^2}{dt^2}q^{(m)}(t) + \omega_m^2 q^{(m)}(t) = -\frac{1}{\epsilon_0}\frac{d}{dt}\left(\int_V \mathbf{J}\cdot\mathbf{e}^{(m)} dV\right) \quad (6)$$

by using the relation $\nabla\times\mathbf{h}^{(m)} = \omega_m/c\,\mathbf{e}^{(m)}$. Here $\mathbf{J}$ is the current source, $\omega_m$ is the angular frequency of mode $m$, $c$ is the speed of light, $\epsilon_0$ is the permittivity of free space, and the integration is performed over the cavity volume. Note that so far there is no restriction on the trajectory and the distribution of the current source $\mathbf{J}$. Assuming a pointlike charge $Q$ to move on a trajectory $\mathbf{x}(t)$, Eq. (6) can be rewritten as

$$\frac{d^2}{dt^2}q^{(m)}(t) + \omega_m^2 q^{(m)}(t) = -\frac{Qc}{\epsilon_0}\frac{d}{dt}[\mathbf{e}^{(m)}(\mathbf{x}(t))\cdot\hat{\mathbf{x}}(\mathbf{t})] \quad (7)$$

with $\hat{\mathbf{x}}(\mathbf{t})$ being the unit vector in the direction of the trajectory. The field distribution $\mathbf{e}^{(TE_{111}\text{-}6)}$ is obtained via CST® [17], and Eq. (7) is solved numerically for different beam trajectories $\mathbf{x}(t)$. In the first scenario, an ultrarelativistic ($|\dot{\mathbf{x}}| = c$) paraxial passing beam with different offsets $x$ is considered. In the second case, the beam traverses the cavity with different tilt angles $x'$ through the longitudinal center of the cavity, as illustrated in Fig. 1. The maximum value of the amplitude $q^{(TE_{111}\text{-}6)}(t)$ is calculated for different beam trajectories. Results are shown in Fig. 4 for different trajectory offsets (left) and angles (right). A linear fit reveals that an amplitude excited by a tilt angle of $x'_0 = 1$ mrad corresponds to an amplitude excited by a trajectory offset of $x'_0 = 214$ $\mu$m. The discrepancy with the estimations made in Ref. [15] is reasonable, since the considered geometry differs. Measuring this ratio is one of the purposes of this paper.

## IV. EXPERIMENTAL SETUP

Figure 5 shows a schematic drawing of the experimental setup used for the HOM-based cavity misalignment measurements. The trajectory of the beam is varied with two pairs of steerers. The beam transverse position is measured by two beam position monitors (BPMs), upstream and downstream of the considered module, respectively. For each trajectory, the dipole mode signal is measured. The basic principle of the involved electronics [18] is shown in the lower part of Fig. 5. A bandpass filter is used to select the $TE_{111}$-6 mode around 1.7 GHz. It is then down mixed with a local oscillator signal at 1.68 GHz to an approximately 20 MHz IF and digitized with 108.3 MS/s. The $TE_{111}$-6 mode frequencies for the different cavities range approximately ±10 MHz around 1.7 GHz, resulting in an IF range from 10 to 30 MHz. The decay times of the HOM

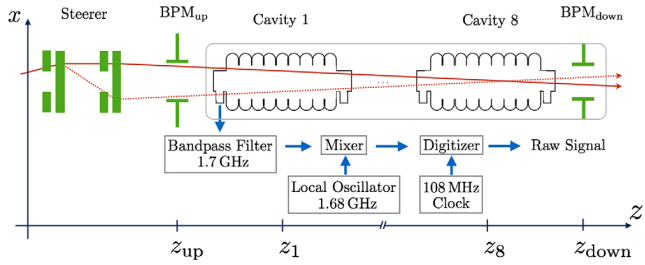

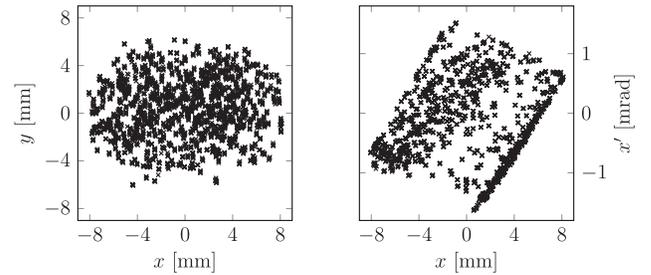

FIG. 5. Schematic drawing of the experimental setup used for the HOM-based cavity misalignment measurement. Eight cavities are located inside one accelerating module, followed by a BPM. The rf is switched off, assuring a drift space between the upstream and downstream BPMs. The beam position and angle at each cavity are calculated from the BPM readings. For each dimension, two steerers are needed in order to get any combination of beam offset and angle at a cavity. Each cavity is equipped with an upstream and a downstream HOM coupler. At each coupler, the signals are filtered, down-converted, and digitized.

FIG. 6. Beam position and angle as calculated for the first cavity of ACC2 at FLASH during a HOM-based cavity misalignment measurement. The $x$-$y$ plane (left) and the $x$-$x'$ plane (right) for 3300 trajectories are plotted. For each plane, two steerers are driven randomly within a defined range, giving, e.g., in the $x$-$x'$-plane a tilted rectangular shape. Its inhomogeneous filling is due to beam losses.

modes vary for each cavity but are typically on the order of microseconds.

In order to relate the measured signal from a HOM coupler with the beam position in that cavity, the beam trajectory has to be known. For a high initial beam energy of several hundred MeV, the transfer matrices of the cavities can be estimated from the rf parameters, and the difference between the actual trajectory and that of a drift space is small. For the first module, however, the difference is significant, and its particular value depends strongly on the beam input parameters. Furthermore, the beam trajectory within one cavity differs significantly from a straight line. A precise knowledge of the beam position and angle in each cavity therefore requires the rf to be switched off. In addition, all quadrupole magnets between the BPMs are cycled to zero field to avoid kicks.

If a drift space can be modeled between the upstream and downstream BPMs, the beam position and angle at the center of each cavity can be interpolated from the BPM readings. The reference axis is defined by these two BPMs. It is worth noting that the downstream BPM is located inside the accelerating module. As a consequence, the measurable offset of the downstream cavities may be underestimated, as will be reviewed later in Sec. VI. The resolution of the involved BPMs is measured before each HOM measurement and is found between 5 and 60 $\mu$m.

As illustrated in Fig. 5, for each transverse plane two steerers are required in order to change the trajectory offset and angle at each cavity to cover a sufficient range for later analysis while obeying machine constraints. The phase space coverage of typical beam trajectories is shown exemplarily for the first cavity of the second accelerating module (ACC2) in Fig. 6. An example of a HOM spectrum from a spare TESLA cavity recorded with a network analyzer is shown in the left part in Fig. 7. Several bands are noticeable, e.g., between 1600 and 1800 MHz ($TE_{111}$ modes), between 1800 and 1900 MHz ($TM_{110}$ modes), and above 2400 MHz ($TM_{011}$ modes). The $TE_{111}$-6 mode around 1680 MHz is highlighted and magnified in the right plot. One notices that the two polarizations from the same $TE_{111}$-6 mode have different frequencies mainly due to the influence of the couplers. The discrepancy normally is of the order of 500 kHz.

The left-hand side of Fig. 8 shows an example of a digitized beam-excited raw signal from cavity eight at ACC1 in the time domain. Before the signal can be analyzed quantitatively, the following digital filtering is applied. The sinusoidal calibration signal of the electronics has to be removed. This can be done in the time domain by fitting a sinusoidal function to the signal prior to the time at which the beam has entered the cavity. The transient of the beam itself also induces a signal which is removed by cutting the signal at the corresponding position. Another distorting effect can be the saturation of the digitizer, which occurs in case of a strong signal amplitude. In order to quantitatively compare dipole modes excited by different beam trajectories, one has to ensure that the saturated part of the signal is removed in all data samples consistently. The resulting processed signal is plotted in the right graph of Fig. 8. Finally, the signal has to be normalized with the bunch charge. Its measurement, however, is not possible at each cavity but only at the cavity-type BPM at the end of the module. Steering the beam in a wide range can cause partial beam losses which may occur between the cavities. Normalization to the measured charge can therefore overestimate the signal in the first cavities at certain beam positions. Care has therefore been taken so that beam losses were limited. For further analysis, it is convenient to normalize the signal for each coupler; in that way, the different trajectories of one set of data correspond to dipole mode amplitudes between zero and one.

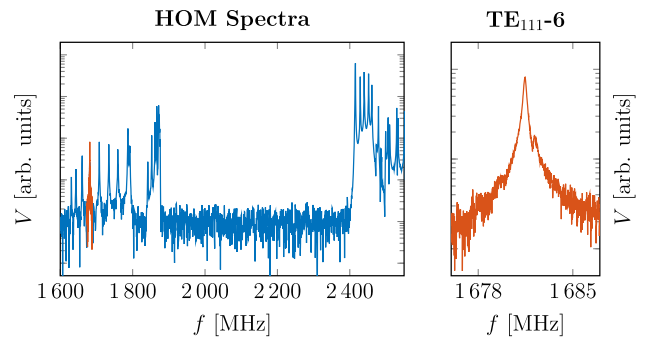

FIG. 7. Normalized beam-excited spectrum as recorded from a HOM coupler at ACC4 at FLASH. The left plot covers a wide frequency range including several monopole and dipole bands. The highlighted peak in the left plot and its magnification in the right one shows the $TE_{111}$-6 dipole mode used for the diagnostic. Note its double-peak character, indicating a frequency deviation between the two parts of the orthogonally polarized doublet.

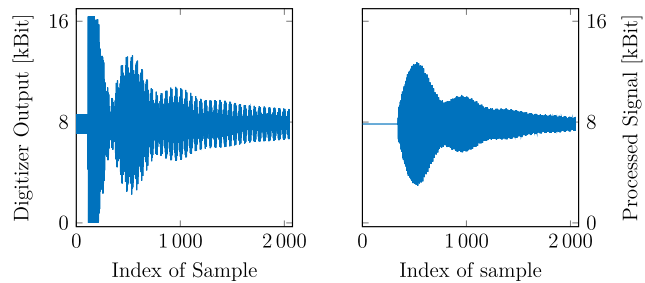

FIG. 8. Beam-excited $TE_{111}$-6 dipole mode in the time domain from the HOM electronics of the downstream coupler of cavity eight in ACC1. The sampling rate is 108 MHz. The measured raw signal (left) and the processed signal (right) are plotted. The oscillation of the raw signal in the first 100 samples is the calibration signal of the readout electronics. For this particular beam trajectory, the digitizer is in saturation till sample 180. Since it is in saturation till sample 343 for other beam trajectories, the processed signal is cut till that point for all trajectories to ensure a quantitative comparison.

## V. ANGULAR DEPENDENCE OF DIPOLE MODES

As described in Sec. II, the field amplitude of the excited dipole mode can be described as a sum of contributions from the beam offset and angle. Restoring each part of the signal individually is hardly possible. If a sufficient number of data points is available, theoretically the linear correlation between each of the beam coordinates $[x, x', y, y']$ and the signal amplitude at each cavity can be resolved. However, as described in the following, this requires a high resolution of the involved devices and is experimentally difficult to achieve. Based on the simulations presented in Sec. III, the scaling factor between the dipole mode amplitude excited by the bunch trajectory offset and the amplitude from the trajectory tilt angle is 214 $\mu$m∶1 mrad. In order to verify this ratio experimentally for a TESLA cavity, a comprehensive measurement procedure was applied at the second cavity of ACC2.

Ideally, the 4D transverse phase space should be filled in order to achieve minimum excitation of the undesired fields from trajectory offsets. However, this requires the coordinated use of four steerers, which is experimentally challenging due to calibration uncertainties. It turned out that subsequent data filtering is more efficient. Therefore, the offset dependence of the two dipole modes is measured in both planes on a sufficient grid of beam trajectories. After postprocessing the raw signal as explained in the previous section, the two amplitudes of the dipole doublet are calculated using a Fourier transformation of the processed signal and a peak-finding algorithm. The polarization axes are found with a fitting routine as illustrated in Fig. 9. A coordinate system transformation $\tilde{x} = x \cos\phi + y \sin\phi$ and $\tilde{y} = y \cos\phi - x \sin\phi$ with the polarization angle $\phi$ ensures that data evaluation is performed on the polarization axes and orthogonal to it, respectively.

The trajectory which minimized the dipole mode was identified through an online optimization routine [19] using upstream steerers and set as a reference trajectory for the further procedure. The ratio between the strength of the two steerers used for each plane in our measurement, as illustrated in Fig. 5, was fixed in order to allow the variation of the trajectory tilt angle without changing the offset in the second cavity of ACC2 substantially.

Multiple data sets taken in different measurement shifts are evaluated. Preliminary results for both polarizations are shown in Figs. 10 and 11, respectively. A total of 23 000 data points and, thus, beam trajectories are evaluated. Every tenth data point is plotted in each graph to reduce the file size. The upper left plot in Fig. 10 shows the amplitude of the second dipole mode (cf. right plot in Fig. 9) measured at the downstream coupler as a function of the transformed horizontal offset $\tilde{x}$ with a polarization angle of $\phi_2 = 86.5°$. The upper right plot shows the same data as a function of transformed horizontal tilt $\tilde{x}'$. The lower two plots show the amplitude as a function of the transformed vertical coordinates $\tilde{y}$ and $\tilde{y}'$, respectively. As expected, the mode amplitude is clearly sensitive only to the transformed vertical plane $[\tilde{y}, \tilde{y}']$. Note that each evaluated data point, and thus beam-excited dipole mode amplitude, is plotted in every graph, thus revealing a complete 4D scan.

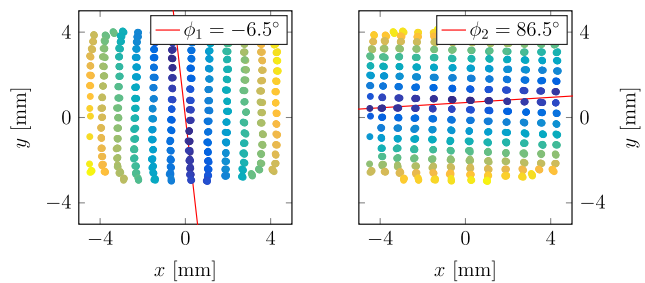

FIG. 9. Normalized amplitude of the first dipole mode at the upstream coupler (left) and of the second mode at the downstream coupler (right) at cavity two in ACC2 as a function of beam offset $x$ and $y$. The dots are color coded by the relative signal strength, giving a bright yellow point at 1 and a dark blue point at 0. The red lines indicate the fitted polarization axes for both modes.

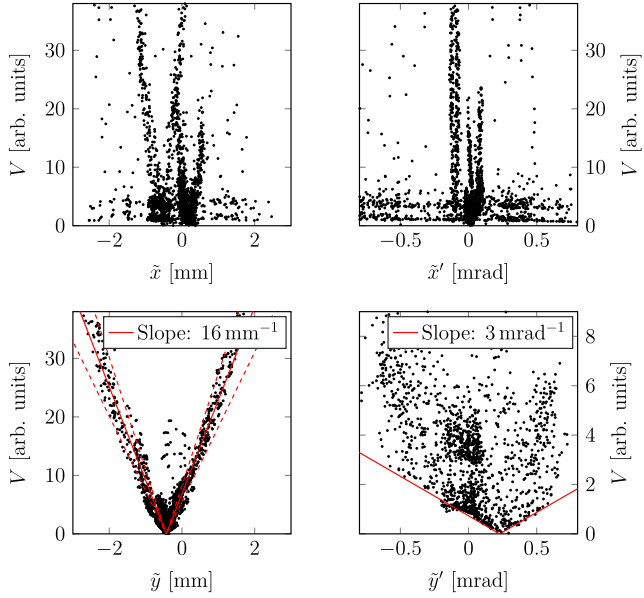

FIG. 10. Normalized amplitude $V$ of the second dipole mode at the upstream coupler at cavity two in ACC2 as a function of the transformed horizontal beam trajectory offset $\tilde{x}$ (upper left), vertical offset $\tilde{y}$ (lower left), horizontal tilt $\tilde{x}'$ (upper right), and vertical tilt $\tilde{y}'$ (lower right). Note that each data point appears in every graph.

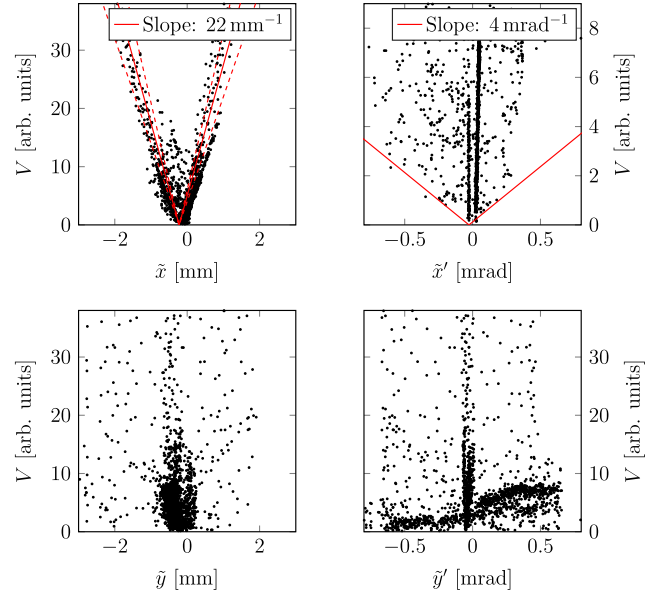

FIG. 11. Normalized amplitude $V$ of the first dipole mode at the downstream coupler at cavity two in ACC2 as a function of the transformed horizontal beam trajectory offset $\tilde{x}$ (upper left), vertical offset $\tilde{y}$ (lower left), horizontal tilt $\tilde{x}'$ (upper right), and vertical tilt $\tilde{y}'$ (lower right). Note that each data point appears in every graph.

It is reasonable to assume that the lowest amplitudes at each $\tilde{y}'$ correspond to a signal purely excited by the beam trajectory tilt with respect to the cavity axis. The overall minimum is reached at $\tilde{y}'_{\min} = 214$ $\mu$rad with respect to the axis defined by the two BPMs. The linear increase of the lowest amplitudes with $\tilde{y}'$ reveals the dipole mode dependence on the trajectory tilt angle and is found to be 3.2/mrad in arbitrary units. Based on the previously mentioned simulations, this corresponds to an offset dependence of 16/mm for the normalized amplitude.

In the lower left graph in Fig. 10, this slope is plotted. The dashed lines indicate a ±20% error. The same analysis was applied for the first dipole mode (cf. left plot in Fig. 9) as measured at the upstream HOM coupler. The result is shown in Fig. 11 in the transformed coordinate system with $\phi_1 = -6.5°$. The minimum amplitude is found at $\tilde{x}'_{\min} = -3$ $\mu$rad.

No well-defined linear correlation between the dipole mode amplitude and the trajectory offset is observed. For example, the linear correlation should result in a well-defined minimum, which is not the case. Systematic errors at the involved BPMs are supposed to be the dominant source of error. Because of the large amount of required data, and thus beam time, the plotted data were measured during three individual shifts, while a precise BPM calibration was performed once. Numerical studies have shown that BPM calibration errors of ±10% can affect the calculated trajectory offset in this particular experimental setup by about ±20% while hardly changing the calculated trajectory tilt. Long-term drifts in the involved readout electronics cannot be excluded and could therefore induce a pseudocorrelation, for example, between $x$ and $x'$. The offset dependence of the dipole mode amplitude of different data sets is slightly different. This supports the assumption that variable BPM calibration errors are affecting the measurement. Furthermore, cavity movement between the measurements cannot be excluded. Future studies may focus on this issue.

However, we interpret these preliminary results as a clear indication that the simulated ratio between the offset and angle dependence of the dipole mode in a TESLA cavity of about 1 mm:5 mrad is appropriate.

## VI. CAVITY MISALIGNMENT MEASUREMENT

Considering the large amount of required data, and thus beam time, it was not possible to apply the previously described measurement procedure to other cavities. Typical ranges for the trajectories during a HOM-based cavity misalignment measurement are $u_{\max} \approx \pm 10$ mm and $u'_{\max} \approx \pm 1$ mrad, where $u$ stands for $x$ and $y$. Based on the previous considerations, it is expected that the maximum signal amplitude related to trajectory offsets exceeds the maximum signal related to tilts by a factor of 50. We will show in the following that during ordinary cavity misalignment measurements the dipole mode excitation due to trajectory tilts can be neglected. Figure 12 illustrates an example of the normalized dipole mode amplitude versus beam offset. In the left plot, the amplitude is color coded, showing qualitatively the dependency. The right graph shows the amplitude of the highlighted points as a function of transformed vertical offset $\tilde{y}$, including a linear fit. The corresponding polarization angle is found as $\phi = 78.5°$. In this example, the offset dependency of the normalized amplitude of the excited mode is about 15%/mm. The phase space coverage is $y_{max} \approx \pm 12$ mm and $y'_{max} \approx \pm 1.2$ mrad. This leads to a maximum amount of less than 2% of the signal amplitude excited by the trajectory angle. Figure 13 shows the residuals of the linear fit shown previously. On the left-hand side, the absolute difference between the normalized amplitude and the linear fit is plotted as a function of the transformed vertical trajectory offset $\tilde{y}$. The right plot shows the same residuals as a function of the transformed vertical trajectory angle $\tilde{y}'$, indicating that there is no distinct systematic identifiable in either graph.

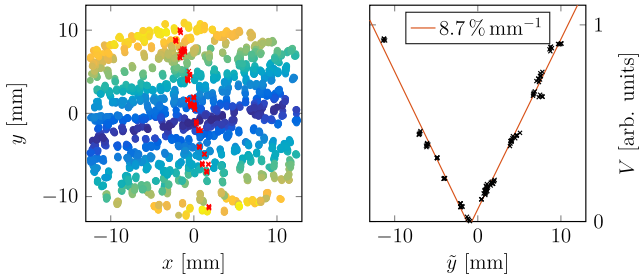

FIG. 12. Normalized amplitude $V$ of the dipole mode at the downstream coupler of cavity eight in ACC2 as a function of beam offset $x$ and $y$. The dots in the left plot show qualitatively the offset dependency and are color coded by the relative signal strength, giving a bright yellow point at 1 and a dark blue point at 0. The normalized amplitude of the highlighted points is plotted in the right graph as a function of the transformed vertical offset $\tilde{y}$, including a linear fit with a slope of 8.7%/mm.

It is nontrivial to obtain the angular dependence from this data set. Based on the previous discussion in Sec. V, it is expected that the angular dipole mode is weakly excited in comparison to the excitation due to trajectory offsets. Thus, during usual cavity misalignment measurements it was not possible to resolve the angular dependency of the dipole mode. As a result, the cavity tilt cannot be determined. In the following discussion, the beam trajectory angle is neglected and serves as a source of uncertainty for the cavity misalignment measurement. However, the phase difference as indicated by Eqs. (1) and (2) might be exploited by the system now under development for the European XFEL. Regarding offset measurements, the experimentally challenging phase space coverage as shown in Fig. 6 is of minor importance. Especially for the injector module, this is a fortunate situation. Machine restrictions such as few available steerers with a limited range and the small aperture thus far prevented a relatively uncorrelated phase space coverage.

As discussed before, the two HOM couplers have different sensitivities to the two polarizations of the same dipole mode. The intersection of the transverse axes where each of these polarizations have minimum amplitude defines the cavity center, actually the center of this particular mode. By identifying the beam position $[x_0, y_0]$ with a minimum dipole power, the electrical center of the mode can be determined.

As illustrated in Fig. 12, the beam phase space can be binned into narrow slices, and a linear function can be fit to the amplitude, providing the center for each slice. These centers provide the axis of each polarization as shown in Fig. 9. However, it turned out that this method was rather difficult to automize. In order to deal with large amounts of experimental data with unequal offset limits for different couplers, a reliable fit algorithm was established. A parabolic fit to the signal power is equivalent to a linear fit to its amplitude. The implemented algorithm fits a 2D parabola of the form

$$f(\tilde{x}, \tilde{y}) = A_{\tilde{x}}(\tilde{x} - \tilde{x_0})^2 + A_{\tilde{y}}(\tilde{y} - \tilde{y_0})^2 \quad (8)$$

to the total power of the signal of each coupler. The rotational transformation

$$\tilde{x} = x \cos\phi + y \sin\phi,$$
$$\tilde{y} = y \cos\phi - x \sin\phi$$

of the beam positions $x$ and $y$ takes into account that the polarization axes of the dipole doublet may vary from the machine transverse axes. The amplitudes $A_{\tilde{x}/\tilde{y}}$, offsets $[\tilde{x}_0, \tilde{y}_0]$, and the skew angle $\phi$ are fit parameters. The fit is done for each coupler individually. A back-transformation $[\tilde{x}_0, \tilde{y}_0] \rightarrow [x_0, y_0]$ provides the position of the minimum in the reference frame as defined by the BPMs. A weighted mean of the individual coupler minima is then used to identify the transverse position of the cavity center. Note that the two polarization axes are assumed to span an angle of 90°. Although this might induce additional error, it turned out that this uncertainty is small compared to the benefit of reduced fit parameters and, thus, robustness of the algorithm. Figure 14 shows a qualitative illustration of the power of the measured dipole mode signal for each coupler at the injector module at FLASH. The missing points at cavity three are caused by momentary problems of the readout electronics. The fitted cavity centers are highlighted as red dots. Note the unequal axes limits for the first four and the last four cavities. It was not possible to fill the $[x, y]$ phase space at the first cavities in a wider range. The rising edge of the dipole mode in the positive vertical direction could not be reached in the first cavities. In addition, it was not possible to switch off the third horizontal gun steerer and still get sufficient transmission through the module. These points hint to a significant horizontal misalignment of the injector module with respect to the gun section.

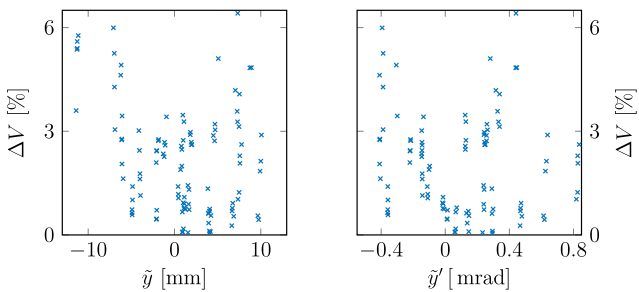

FIG. 13. Dipole mode residuals $\Delta V$ of the linear fit presented in Fig. 12 as a function of the transformed vertical trajectory offset $\tilde{y}$ (left) and transformed vertical trajectory angle $\tilde{y}'$ (right). No distinct systematic is identifiable in either graph.

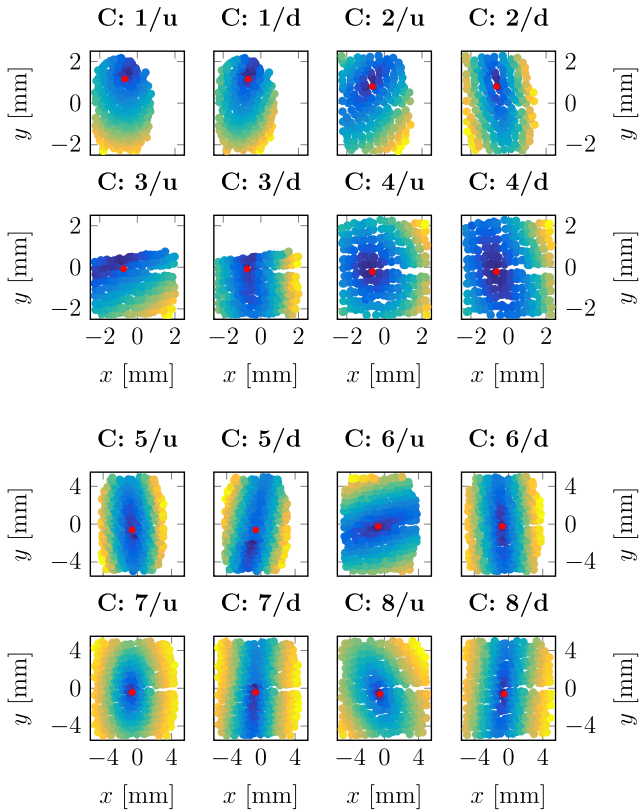

FIG. 14. Normalized power of the dipole mode as measured at each coupler at the injector module at FLASH as a function of beam offset $x$ and $y$. The upper left plot shows the upstream coupler of the first cavity, $C:1/u$, for example, and the lower right plot the downstream coupler of the eighth cavity, $C:8/d$. The dots are color coded by the relative signal strength, giving a bright yellow point at 1 and a dark blue point at 0. The red dots indicate the fitted cavity centers. Note the unequal axes limits for cavities 1–4 and 5–8.

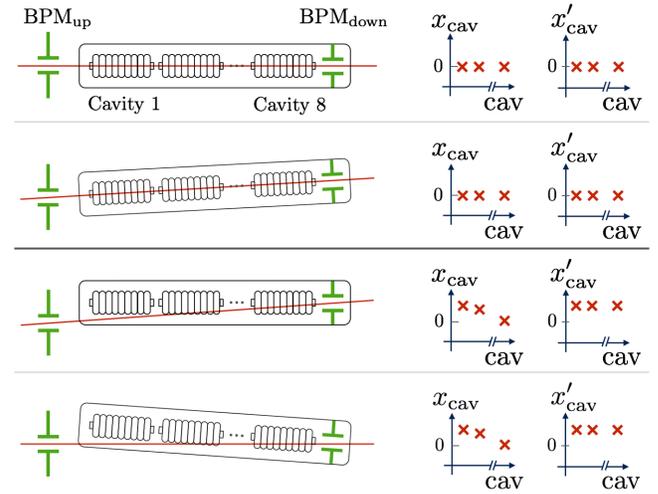

FIG. 15. Schematic drawing of the HOM-based cavity misalignment measurement setup as described in Fig. 5. The drawing illustrates exemplarily four module alignment scenarios (left) for the horizontal plane. The right-hand side shows the corresponding measurement result. The cavity offsets $x_{cav}$ and tilts $x'_{cav}$ are measured with respect to the axis which is defined by the upstream and downstream BPM (red line). The latter BPM is located inside the cryomodule. The upper two and the lower two cases, respectively, are indistinguishable.

As mentioned before, the used reference axis is defined by the zero readings of the BPMs. Because of experimental constraints, the required downstream BPM available for HOM-based misalignment measurements is located behind the last cavity, still inside the accelerating module. Figure 15 illustrates the thereby arising difficulties. Illustrated are different alignment scenarios of the injector module (left) and corresponding measurement results for the cavity misalignments (right) with respect to the reference axis.

Even if both the offset and angle of each cavity would be measurable, it would still not be possible to distinguish, for example, between a perfectly aligned module and a module which is rotated around the upstream BPM. As the lower two examples point out, the same is true for any setup rotated around the upstream BPM. Moreover, the measurable offset of cavity $n$ with respect to the design axis of the machine will decrease linearly with $n$. Despite these uncertainties, a quantitative analysis of the cavity misalignment measurement follows.

Figure 16 shows the transverse electrical cavity centers obtained with our method at the injector module at FLASH as obtained from two data sets, separated by about three months of machine operation. The fit results (cf. Fig. 14) in the reference frame as discussed above are plotted in the upper row. It is reasonable to define the module axis by the individual cavity centers. A linear fit therefore reveals the module axis with respect to the reference frame. The lower row shows the cavity misalignment with respect to the module axis. The agreement between the two individual measurements is satisfactory, considering the described experimental challenges and physical movement of the cavities. Combining both measurements, the fitted module axis can be described to have an offset and tilt with respect to the gun section of

$$[\Delta x_{\text{mod}}, \Delta y_{\text{mod}}] = [(-0.65 \pm 0.21)~\text{mm}, (0.03 \pm 0.2)~\text{mm}],$$
$$[\Delta x'_{\text{mod}}, \Delta y'_{\text{mod}}] = [(22 \pm 12)~\mu\text{rad}, (-229 \pm 6)~\mu\text{rad}].$$

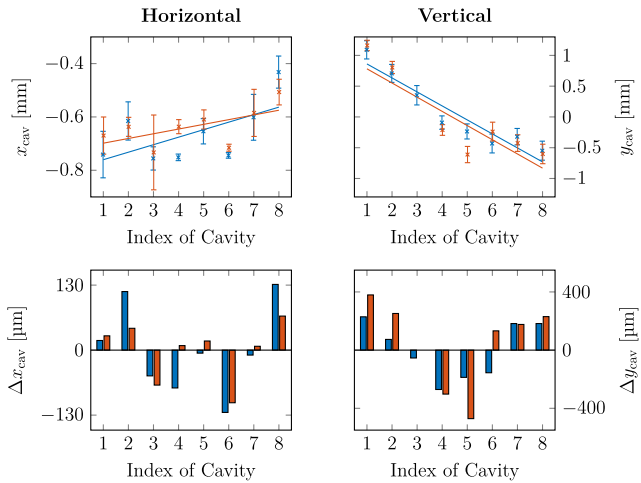

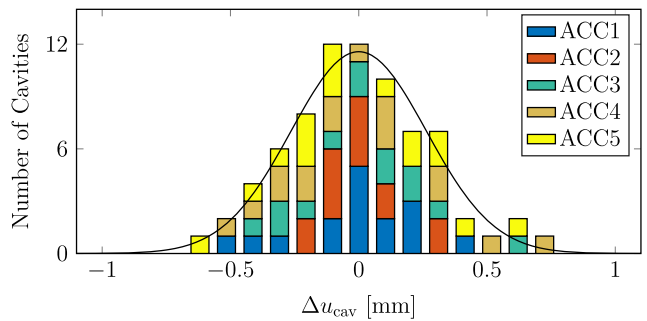

FIG. 16. HOM-based cavity misalignment measurements for the horizontal (left) and vertical (right) planes. The results of two independent measurement shifts are plotted in red and blue, respectively. The upper plot shows the calculated cavity center $u_{\text{cav}}$ (dots) and the deduced module axis (lines) with respect to the BPM reference axis. The middle row illustrates the residual cavity offset $\Delta u_{\text{cav}}$ with respect to the calculated module axis.

FIG. 17. Histogram of the fitted residual cavity offset $\Delta u_{\text{cav}}$ with respect to the calculated module axis of the first five accelerating modules at FLASH. Both transverse planes are included. The fitted standard deviation is $\sigma_{\text{cav}} = 342~\mu$m.

However, the quantitative interpretation of these values is rather difficult. Results shown in Fig. 14 must be taken into account as well as the above-mentioned problems of transmission through the module, when operating gun steerers below their maximum value. The fact that all horizontal steerers had to be set near their maximum value suggests that the relatively small fitted values $[\Delta x_{\text{mod}}, \Delta x'_{\text{mod}}]$ are underestimating the actual misalignment (see, for example, the second row in Fig. 15). Furthermore, recall the insensitivity of the measurable offset of downstream cavities caused by the used reference axis. The fitted values of vertical offsets do not reflect that prediction. A likely explanation is that the involved downstream BPM shows a significant error. In order to undeniably determine the actual misalignment of the injector module with respect to the gun section, additional measurement efforts have to be made.

The principle of the above-described procedure for the measurement and data analysis was applied also to the next four accelerating modules (ACC2–5) at FLASH, which are equipped with HOM readout electronics. Because of the higher beam energy at downstream modules, and thus smaller beam size and lower sensitivity to space charge effects, we were able to implement an efficient measurement procedure while obeying the experimental constraints.

In contrast to the injector module, the [x, y] phase space could be filled reasonably well. A total of 32 000 beam trajectories are evaluated. A histogram of the residual cavity offsets as obtained from the corresponding fitted module axes is shown in Fig. 17, combining both transverse planes. A Gaussian distribution reveals a standard deviation of $\sigma_{\text{cav}} = 342~\mu$m. The measurement result agrees decently with the specification of the maximum cavity offset of $\Delta u_{\text{cav}} = \pm 500~\mu$m in TESLA modules [7]. The exact values of previous measurements of the cavity offsets in ACC4 and ACC5 made in 2006 [10] were not reproduced. However, this is expected, since the modules have been heated up between the measurements, which in turn changes the cavity alignment significantly [20].

## VII. SUMMARY

Beam-excited dipole modes which are extracted by HOM couplers can be used to study the transverse position of the accelerating cavities inside an installed cryomodule.

We have developed and applied an efficient measurement and data evaluation procedure on five accelerating modules at FLASH, which are equipped with HOM readout electronics. The rms value of the measured offset of the cavities in the first five accelerating modules of 342 $\mu$m agrees with the specification of maximum cavity offset of 500 $\mu$m in TESLA modules.

We performed eigenmode simulations of the TESLA cavity showing that the ratio between the offset and angle dependence of the dipole mode is about 1 mm:5 mrad. For the first time, we measured the dipole mode excitation due to trajectory tilt angles. The preliminary results agree with the simulated values.

## ACKNOWLEDGMENTS

This research was supported by the German Academic Scholarship Foundation and by EuCARD-2, European Commission Grant Agreement No. 312453. We thank Micha Dehler for supporting the measurement and fruitful discussions. We extend our thanks to Martin Dohlus for valuable help with the eigenmode simulations.